# An RSSI-based Wireless Sensor Node Localisation using Trilateration and Multilateration Methods for Outdoor Environment

**Mohd Ismifaizul Mohd Ismail[1], Rudzidatul Akmam Dzyauddin[2], Shafiqa Samsul[3], Nur Aisyah Azmi[4], Yoshihide Yamada[5], Mohd Fitri Mohd Yakub[6], Noor Azurati Binti Ahmad @Salleh[7]**

[1,2,7] Razak Faculty of Technology and Informatics, Universiti Teknologi Malaysia Kuala Lumpur, Jalan Sultan Yahya Petra, 54100 Kuala Lumpur, Malaysia.
[1]ismifaizul@gmail.com, [2]rudzidatul.kl@utm.my, [7]azurati@utm.my
[4,5,6]Department of Electronic System Engineering (ESE), Malaysia - Japan International Institute of Technology, University Teknologi Malaysia Kuala Lumpur, Jalan Sultan Yahya Petra, 54100 Kuala Lumpur, Malaysia,
[3]shafiqasamsul@gmail.com, [4]aisyah.azmi@yahoo.com, [5]yoshihide@utm.my, [6]mfitri.kl@utm.my,

**ABSTRACT**

Localisation can be defined as estimating or finding a position of node. There are two techniques in localisation, which are range-based and range-free techniques. This paper focusses on Received Signal Strength Indicator (RSSI) localisation method, which is categorised in a range-based technique along with the time of arrival, time difference of arrival and angle of arrival. Therefore, this study aims to compare the trilateration and multilateration method for RSSI-based technique for localising the transmitted (Tx) node. The wireless sensor module in the work used LOng-RAnge radio (LoRa) with 868 MHz frequency. Nowadays, wireless networks have been a key technology for smart environments, monitoring, and object tracking due to a low power consumption with long-range connectivity. The number of received (Rx) nodes are three and four for trilateration and multilateration methods, respectively. The transmitted node is placed at 32 different coordinates within the 10x10 meter outdoor area. The results show that error localisation obtained for General Error Localisation (GER) for multilateration and trilateration is 1.83m and 2.30m, respectively. An additional, the maximum and minimum error for multilateration and trilateration from 1.00 to 5.28m and 0.5 to 3.61m. The study concludes that the multilateration method more accurate than the trilateration. Therefore, with the increase number of Rx node, the accuracy of localisation of the Tx node increases.

## 1. INTRODUCTION

Localisation study has been a trend in finding and estimating a node [1], [2]. Global Positioning System (GPS) is one of the straight forward methods to obtain localisation position, but requires a direct line of sight (LOS). Nevertheless, it seems unfeasible to be implemented for shadowing environment [3]–[5]. Various applications [6], [7] such as missile guidance systems, habitat monitoring, medical diagnostics and objects tracking employed wireless sensor localisation.

Wireless sensor network (WSN) is the sensor nodes that transmitting the wireless signal data to the administration centre or control centre from different location within a specific area. The sensor in WSN used infrastructure-based or a base station to receive the data from a specific sensor node at a specified location. All the information and particular data aim at their environment using sensor nodes within the area [8]. The WSN is an ad-hoc wireless network and some applications are hard to reach or at hazardous areas as illustrated Table 1. The WSN has comprised of hundreds to thousands of nodes with inhibited computing power and limited memory, where the short battery life will give the main obstacle in distance estimation of nodes. Localisation is a significant issue for various sensor applications, and the accuracy of distance estimation should be higher, and localisation sensor must be low cost. Localisation study has been a trend in finding and estimating a node [5], [17], [18]. GPS is one of the straight forward methods to obtain the localisation position, but requires a direct line of sight (LOS). It seems unfeasible to be implemented for shadowing environment [6]. Various applications [6] such as missile guidance systems, habitat monitoring, medical diagnostics and objects were tracking employed wireless sensor localisation.

**Table 1:** Wireless sensor network application

| Industries | Application |
|---|---|
| Enforcement (military, police, immigration etc) [9]–[12] | • Monitoring<br>• Tracking<br>• Security<br>• Control<br>• Maintenance |

| Industry [13], [14] | • Machine monitoring<br>• Particularly in an area hard to reach |
|---|---|
| Aviation | • Replacing wires networks<br>• Already implement |
| Environment [15] | • Environment monitoring in building<br>• Oceans<br>• Forests<br>• River<br>• Lake<br>• Animal tracking |
| Traffic | • Road control and monitoring<br>• Parking lots<br>• Road LED signboard |
| Engineering [16] | • Monitoring (and modelling) structures |

Figure 1 shows the localisation methods taxonomy. In localisation for wireless sensor networks, target localisation and node self-localisation play a vital role in a node positioning. In target localisation, we can classify into two techniques: (i) single target localisation (ii) multiple target localisation. Target localisation estimates the target position from multiple noisy sensors measurement [20], [21]. On the other hand, a node self-localisation can be broadly categorised into range-free and range-based methods. Range-free is a method to calculate the node position using the distance between the transmitter node to receiver nodes, which the methods are namely centroid, DV-hop and geometry conjecture techniques. The other node self-localisation is range-based includes several techniques: (i) received signal strength indicator (RSSI), (ii) angle of arrival (AoA), (iii) time of arrival (ToA), and (iv) time difference of arrival (TDoA). The advantages of node self-localisation are cost-efficient, quickly deployed and low power consumption [19]. Although self-node localisation methods have high accuracy, an additional hardware is required.

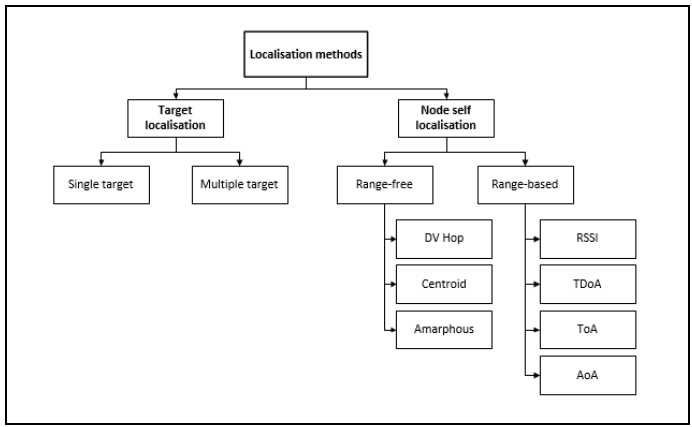

**Figure 1:** Localisation methods taxonomy [2]

For previous localisation works, the problem of the target node was increasing attention due to their applications. The three problems are due to the outdoor environment, low accuracy level and high-cost network technology. The RSSI-based technique with a trilateration algorithm has been proposed for the indoor building by using WiFi and achieved a high node accuracy, but has a minor problem in the study [5], [17]. The problem is that low WiFi access in the indoor building to localise the node and thus defective algorithm is applied. Labyad *et al.* [6] proved that the multilateration algorithm is more accurate than the trilateration algorithm by using the same technique. Sharmilla Mohapatra *et al.* [19] highlighted the approaches of AoA and provided an overview of other techniques. The work also mentioned the high-cost of the antenna array which required by AoA technique in localisation. In order to overcome the high-cost problem, RSSI-based technique is applied for the localisation [19].

Table 2 shows the comparison of range-based and range-free techniques used in sensor node localisation. The table shows the advantage of the RSSI method does not require specialised hardware and low in cost. However, attenuation occurs. Furthermore, AoA shows that using a high cost due to antenna array is needed in this technique [19].

**Table 2:** Comparison of WSN Localisation Methods

| | Principal of operation | Special Hardware | Attenuation problem | Cost |
|---|---|---|---|---|
| **Range-based techniques** | | | | |
| **RSSI** [2], [4] | Signal strength measurement | Not required | High | Low |
| **AoA** [15] | The angle of signal arrival | Required | Medium | High |
| **ToA** [15] | Time of arrival | Required | Low | Med |
| **TDoA** [16] | The time difference in propagation at different points | Required | Low | Med |
| **Range-free technique** [17] | | | | |
| **DV Hop** | The heterogeneous network that consists of sensing nodes and anchors | Not required | High | Low |
| **Centroid** | Use transmit beacons containing (Xi, Yj) | Not required | High | Low |
| **Amorphous** | Takes a different approach from DV-Hop for average single hop | Not required | High | Low |

On the other hand, ToA and TDoA techniques have medium cost and low attenuation problem despite that both require specialised hardware in the sensor node localisation method.

In this paper, the contributions are as follows
- A development of the wireless sensor node prototype for localisation experimental for outdoor without GPS receiver.
- Performance analysis and comparison for different types of the trilateration and multilateration method for Tx node localisation using RSSI data from Rx node.

This paper is organised into six sections. Sensor node localisation techniques are reviewed in Section 2. Section 3 discusses the system design and architecture for GPS localisation sensor node using LoRa transceiver module. Then, the experiment setup for data collection theoretical and location data is explained in the Section 4. Section 5 discusses the result and analysis from multilateration and trilateration method to estimate the Tx node location coordinate using three and four Rx nodes. In additional, the location error (ER) and the general location error (GER) are discussed in the finding. The conclusion and recommendations for future works are explained in Section 6.

## 2. PREVIOUS WORKs

Received Signal Strength Indicator (RSSI) is one of the range-based techniques in localisation. RSSI is categorised as a range-based technique because it is based on a signal attenuation. The longer distance resulted in the signals travel with greater attenuation. Usually, it uses trilateration algorithm to find the position or location of an anchor or target node. RSSI techniques do not require external or additional hardware to perform localisation. Thus, it becomes less costly. However, there are some errors during the distance estimation due to the inconsistency in Radio Frequency (RF) signal propagation. The other factors that influence the RSSI calculation are multipath fading, shadowing effects and attenuation of signals [5], [17]. A significant shadowing errors is found when there are obstacles present.

Figure 2 illustrates the trilateration and multilateration method with the RSSI signal from the receiver node (Rx). The transmitter node and receiver node are employed to determine the position of a target node. In trilateration, at least three transmitters are required to estimate the position. The intersection of three circles around the beacon yields a point that indicates the position of the transmitter node [4], [5], [17].

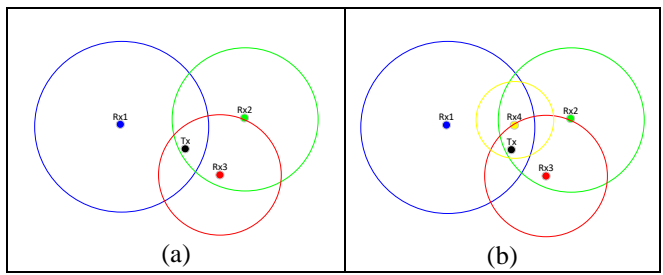

**Figure 2:** RSSI in localization using (a) trilateration and (b) multilateration technique

Table 3 compares the RSSI-based algorithms for WSN localisation system. RSSI has several algorithms: (i) Min-Max, (ii) Multilateration, (iii) Maximum Likelihood, (iv)Ring Overlapping based on Comparison of Received Signal Strenght Indicator (ROCRSSI). Min-Max is one of the RSSI-based algorithms, and it offers the simplicity of implementation. The accuracy of the Min-Max algorithm is dependent on the intersection area in which the smaller area increases the accuracy of the sensor node location. On the contrary, the multilateration algorithm is simple in detecting the sensor node. However, the algorithm is slightly complex compared with the Min-Max algorithm. The benefit of the multilateration algorithm is that high performance can be obtained. The Maximum likelihood is rather complicated compared with the multiliteration. However, the algorithm minimises the variance of error estimation. Finally, ROCRSSI algorithm has high complexity as a trade-off to good performance achieved.

**Table 3:** RSSI-based algorithms comparison

| Algorithm | Complexity | Accuracy | Error |
|---|---|---|---|
| Min-Max | Low | Low | Low |
| Trilateration | Medium | Medium | High |
| Multilateration | Medium | Medium | High |
| Maximum Likelihood | High | High | Low |
| ROCRSSI | High | Medium | Medium |

## 3. SYSTEM DESIGN AND ARCHITECTURE

Figure 3 shows a block diagram of the Wireless Sensor Node GPS tracker system without energy harvesting. Two main components of the wireless sensor node GPS tracker system such as transmitter and receiver nodes. The transmitter node (Tx) designed with four components, which are Global Positioning System (GPS) receiver, Real Time Clock (RTC), Microcontroller Unit (MCU) and LoRa transceiver module. However, the receiver node (Rx) has designed with the same component of the transmitter with an additional Secure Digital (SD) card for storing the data from Tx and Rx. The prototype is developed for experiments of sensor node positioning using the RSSI under the trilateration and multilateration techniques.

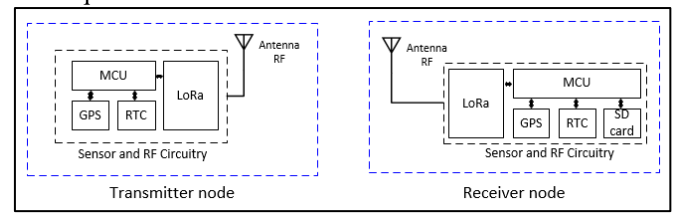

**Figure 3:** A block diagram of the wireless sensor positioning system

Figure 4 illustrates the Tx and Rx sensor node hardware development for localisation data collection. The Rx sensor node is developed using LoRa development, GPS, SD Card and RTC module. However, the Tx sensor node is developed without SD Card module to reduce the power consumption.

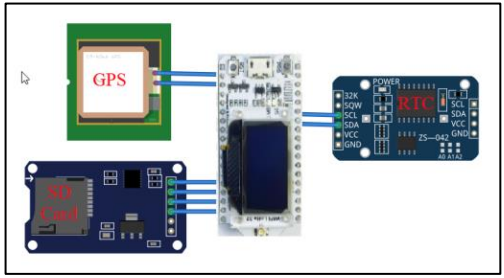

**Figure 4:** Tx and Rx sensor node hardware

## 4. EXPERIMENTAL SETUP

Figure 5 illustrates the experimental setup for the Tx and Rx node localisation. The experiment was conducted according to the procedure at an outdoor area, so that the sensors can directly communicate in line of sight (LOS), which has no obstacles in 10m x 10m area. Every Rx was measured the Received Signal Strength Indication (RSSI) and Signal over Noise Ratio (SNR) from the Tx. In order to find the Tx node location, the path loss model was used to calculate the distance between Rx and Tx nodes. The estimated distance between the Tx to Rx is calculated based on the RSSI value generated at every Rx node.

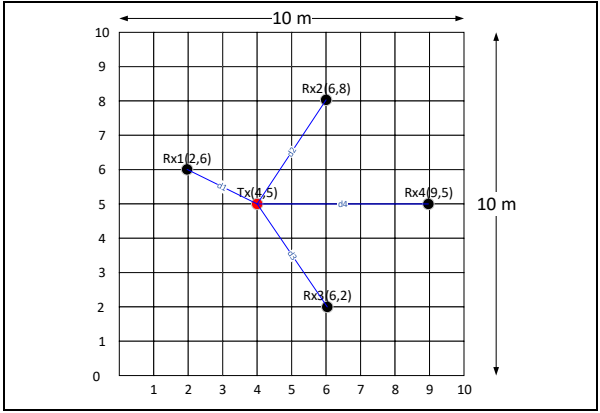

**Figure 5:** Layout node location for the outdoor experiment

The experiment was conducted outdoor environment at the tennis court at University of Technology Malaysia (UTM), as shown in Figure 6. The receiver nodes were placed at position 1, 2, 3 and 4 with the coordinates of (2,6), (6,8), (6,2) and (9,5), respectively (see Figure 5). The target node was initially placed at coordinate (4,6) and changed to 32 different positions — the change of position with a minimum distance of 1 meter to a maximum of 10 meters. The Rx and Tx nodes were set high from the ground approximately 1 and 2 meters because to ensure minimal interference occurs during the signal transmission. The multilateration method was analysed from four Rx node, which the data collected was stored in tge SD card for offline analysis. However, the trilateration method only analysed data from the three Rx noded, which are from Rx1 to Rx3 (see Figure 5).

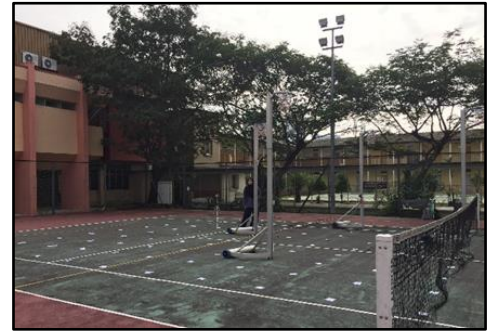

**Figure 6:** Outdoor experiment

The distance between the transmitting and receiving nodes can be calculated with below information,
- Transmitted power of the node
- RSSI
- Path loss model
- Location error (ER) and general location error (GER)

From that three-parameter, the power of the received signal, $P_R^{ij}$ at receiver node, $j$ and transmitter node, $i$ at time, $t$ is expressed as equation (1)

$$P_R^{ij} = P_T^i - 10\eta \log\left(\frac{d_{ij}}{d_o}\right) - x_{ij}(t) \quad (1)$$

Where

$P_R^{ij}$ is power received signal at receiver node, $j$ from transmitter node (Tx), $i$ at the time, $t$. $P_T^i$ is transmitted power at the transmitter node, $i$. $\eta$ is attenuation constant, which is dependent on surrounding of the receiver node, $j$. $d_{ij}$ is the distance between transmitter node, $i$ and receiver node (Rx), $j$. $x_{ij}(t)$ is uncertainly channel model, which the value depends on multipath fading and shadowing. The $d_o$ is reference distance between Tx and Rx node.

From Equation (1), the distance between transmitter and receiver nodes can be estimated using Equation (2).

$$d_{ij} = d_o 10^{\frac{P_T^i - P_R^{ij}}{10\eta}} \quad (2)$$

For different distance, each RSSI value obtained and the path loss exponent, *PLO* and attenuation factor, $\eta$ was calculated. The value for *PLO* is 32.769, whereas $\eta$ is 2.185.

The location for the trilateration method can be estimated using Equation (3).

$$d_{ij}^2 = (x - x_{ij})^2 + (y - y_{ij})^2 \quad (3)$$

where $x$ and $y$ are the estimated location coordinates for Tx node. $x_{ij}$ and $y_{ij}$ are the actual location coordinate of the Rx node. For example, to estimate the location of the Tx node for three Rx nodes (Trilateration) is

$$\begin{cases} (x - x_1)^2 + (y - y_1)^2 = d_1^2 \\ (x - x_2)^2 + (y - y_2)^2 = d_2^2 \\ \vdots \\ (x - x_n)^2 + (y - y_n)^2 = d_n^2 \end{cases} \quad (4)$$

Equation (4) will be linearized as presented in Equation (5) to (7) by subtracting the last equation from the n-1 previous ones.

$$\begin{cases} -2(x_1 - x_n)x - 2(y_1 - y_n)y = (d_1^2 - d_n^2) - (x_1^2 - x_n^2) + (y_1^2 - y_n^2) & (5) \\ -2(x_2 - x_n)x - 2(y_2 - y_n)y = (d_2^2 - d_n^2) - (x_2^2 - x_n^2) + (y_2^2 - y_n^2) & (6) \\ \vdots \\ -2(x_{n-1} - x_n)x - 2(y_{n-1} - y_n)y = (d_{n-1}^2 - d_n^2) - (x_{n-1}^2 - x_n^2) + (y_{n-1}^2 - y_n^2) & (7) \end{cases}$$

Rewrite Equations (4) and (5) to (7) in matrices

$$AX = B \quad (8)$$

Where

$$A = \begin{bmatrix} -2(x_1 - x_n) & -2(y_1 - y_n) \\ -2(x_1 - x_n) & -2(y_1 - y_n) \\ \vdots & \vdots \\ -2(x_{n-1} - x_n) & -2(y_{n-1} - y_n) \end{bmatrix}, X = \begin{bmatrix} x \\ y \end{bmatrix},$$

$$B = \begin{bmatrix} (d_1^2 - d_n^2) - (x_1^2 - x_n^2) - (y_1^2 - y_n^2) \\ (d_2^2 - d_n^2) - (x_2^2 - x_n^2) - (y_2^2 - y_n^2) \\ \vdots \\ (d_{n-1}^2 - d_n^2) - (x_{n-1}^2 - x_n^2) - (y_{n-1}^2 - y_n^2) \end{bmatrix}$$

Then the estimated location coordinate of Tx can be solved using Equation (9)

$$X = A^{-1} B \quad (9)$$

An error distance, $e_{ij}$, is calculated using equation (10)

$$e_{ij} = d_{ij} - d \quad (10)$$

where $d_{ij}$ is an estimated distance and $d$ is an actual distance between Tx and Rx nodes. The Error Rate, *ER* for the Tx coordinate can be computed using Equation (11)

$$ER = \sqrt{(x - x_{ij})^2 + (y - y_{ij})^2} \quad (11)$$

where the $x_{ij}$ and $y_{ij}$ are the estimated coordinate of Tx. Then $x$ and $y$ denoted the actual coordinate of Tx. Equation (12) represents General Error Rate (GER) for the total of the average error between actual and estimated coordinates of Tx.

$$GER = \frac{1}{n}\sum_{i=1}^{n} ER(i) \quad (12)$$

## 5. RESULTS AND ANALYSIS

This section discusses the results obtained from the experiments, which arr RSSI versus distance, estimation of Tx sensor node location and location error (ER) and general location error (GER).

### 5.1. Experimental RSSI vs distance

Figure 7 shows the relationship between RSSI and distance between transmitter and receiver node for outdoor, which is inversely exponential. When the distance between Tx and Rx node is increased, the RSSI value between transmitter and receiver node gradually decreases. The fall of the RSSI is due to the pathloss effect.

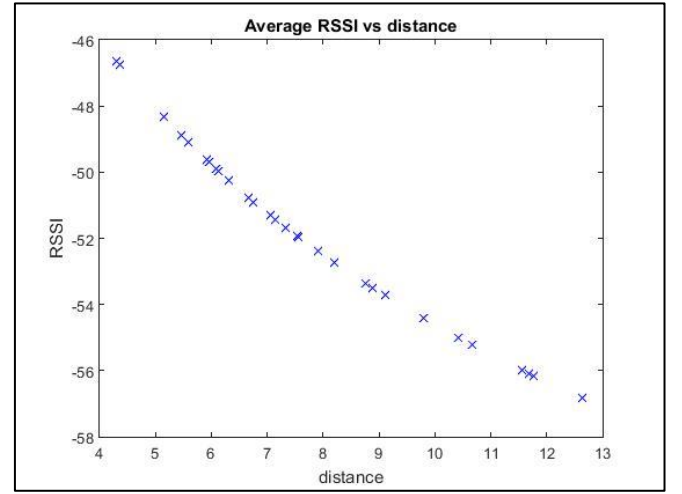

**Figure 7:** Average RSSI vs Distance

### 5.2. Estimation for Tx coordinate

Figure 8 illustrates the estimated coordinate of x and y for Tx node using trilateration and multilateration methods. The trilateration method for actual and estimated location using three Rx nodes, which are Rx1, Rx2 and Rx3. Whereas, the estimated Tx node for the multilateration method using four Rx nodes (such as Rx1, Rx2, Rx3 and Rx4). The figure shows that the estimated locations of Tx for the multilateration are closer to the actual location and several locations are predicted accurately. Overall, an increase of Rx node achieves higher accuracy in detecting the location of the node compared with the trilateration. However, the trade-off is the computation complexity.

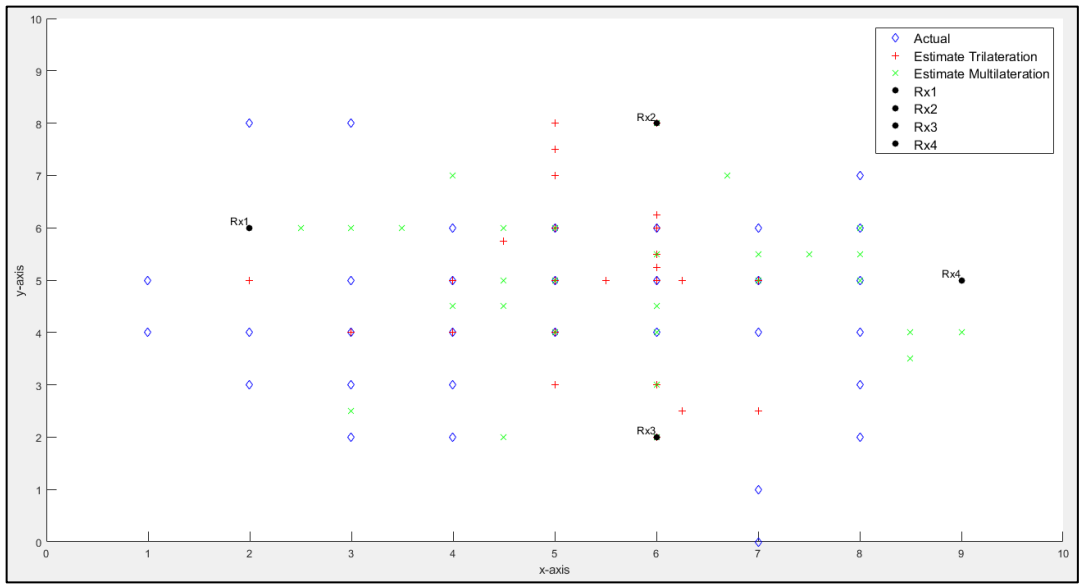

**Figure 8:** Data plot between actual and estimation of Tx node coordinate location; (+) trilateration and (x) multilateration method

*5.3. Location efrror (ER) and general location error (GER)*

A location error of the Tx measured by the multilateration and trilateration methods is shown in Figure 9. The location error shows fluctuation value for both methods. Generally, the GER value decreases with the increasing Rx node, which is demonstrated in the multilateration method. As can be seen, the GER values of the multilateration and trilateration methods are 1.83 and 2.30 meters, respectively.

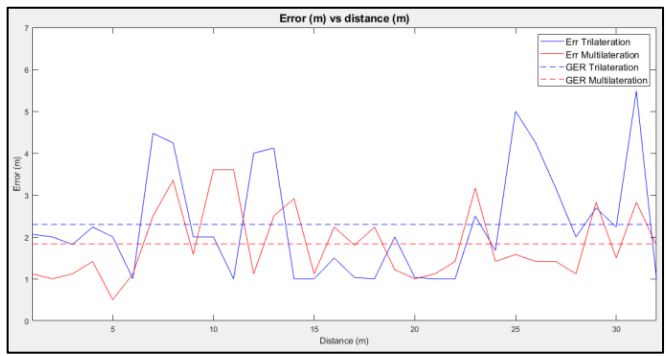

**Figure 9:** Error (m) versus distance (m) for trilateration and multilateration method

## 6. CONCLUSION

In this investigation, it was shown that the RSSI value of the Rx node is gradually decreased when the distance increased. The location of Tx is predicted quite accurately using the multilateration. In fact, the GER value for multilateration is 1.83, which is lower than the trilateration method. It can be concluded that by increasing the number of Rx nodes can increase the accuracy of estimating the Tx location, but the computation complexity arises. Therefore, for future works, the comparison between both techniques can be conducted for indoor.

## ACKNOWLEDGEMENT

This work is supported by the Universiti Teknologi Malaysia under PAS, with cost center no Q.K130000.2740.00K70. We would also like to thank the Ministry of Higher Education and High Center of Excellence Wireless Communication Center (WCC) for funding the publication of the paper under R.K130000.7840.4J235. We would like to extend our gratitude to U-BAN members for comments on the work.